\documentclass{article}
\usepackage{arxiv}
\usepackage[utf8]{inputenc} % allow utf-8 input
\usepackage[T1]{fontenc}    % use 8-bit T1 fonts
\usepackage{hyperref}       % hyperlinks
\usepackage{url}            % simple URL typesetting
\usepackage{booktabs}       % professional-quality tables
\usepackage{amsfonts}       % blackboard math symbols
\usepackage{nicefrac}       % compact symbols for 1/2, etc.
\usepackage{microtype}      % microtypography
\usepackage{lipsum}
\usepackage{graphicx}

\usepackage[sorting=none,url=false]{biblatex} %Imports biblatex package

\usepackage[utf8]{inputenc}
\usepackage{upgreek}
\usepackage{bm}
\usepackage[T1]{fontenc}
\usepackage{mathtools}
\usepackage[thinc]{esdiff}
\usepackage{amssymb}
\usepackage{color}
\usepackage{soul}
\usepackage{appendix}
\usepackage{graphicx}
\usepackage{amsmath}
\usepackage{amsfonts}
\usepackage{amsthm}
\usepackage{setspace}
\usepackage{verbatim}

\usepackage{doi}
\usepackage{bm}
\usepackage{ar}
\usepackage{orcidlink}

\usepackage[justification=justified,singlelinecheck=false,labelsep=endash,position=top]{subcaption}
\usepackage{multirow}
\usepackage{multicol}
\usepackage{etoolbox}

\graphicspath{ {./images/} }

\newcommand{\Pe}{\mbox{P\'e} }

\newcommand{\Peclet}{P\'eclet }
\newcommand{\vK}{{\cal K}}

\newcommand{\Vsp}{V^{\text{s}}{'}}
\newcommand{\Vspsq}{V^{\text{s}}{'}^{2}}

\newcommand{\difft}{\text{d}}
\newcommand{\uvec}{\mathbf{u}}
\newcommand{\xa}{\mathbf{a}}
\newcommand{\xb}{\mathbf{b}}
\newcommand{\xc}{\mathbf{c}}

\newcommand{\xJ}{\mathbf{J}}

\newcommand{\xu}{\mathbf{u}}
\newcommand{\xx}{\mathbf{x}}
\newcommand{\xp}{\mathbf{p}}
\newcommand{\xq}{\mathbf{q}}
\newcommand{\xD}{\mathbf{D}}
\newcommand{\Wo}{\textrm{Wo}}

\newcommand{\Delt}{\Delta t}

\newcommand{\be}{\begin{equation}}
\newcommand{\ee}{\end{equation}}

\newcommand{\dtp}[1]{\frac{\partial #1}{\partial t}}
\newcommand{\dth}[1]{\frac{d #1}{d \theta}}
\newcommand{\dtt}[1]{\frac{d^2 #1}{d \theta^2}}
\newcommand{\cL}{{\cal L}}
\newcommand{\cM}{{\cal M}}
\newcommand{\hG}{\hat{\mathbf{G}}}

\title{Fine-tuning the dispersion of active suspensions with oscillatory flows}

\author{
 Hakan Osman Caldag \\
  Department of Mathematics\\
  University of York\\
  York, UK, YO10 5DD \\
  \texttt{hakan.caldag@york.ac.uk} \\
   \And
 Martin Alan Bees \\
  Department of Mathematics\\
  University of York\\
  York, UK, YO10 5DD \\
  \texttt{martin.bees@york.ac.uk}
}

\begin{document}
\maketitle
\begin{abstract}
The combined impact of axial stretching and cross-stream diffusion on the downstream transport of solute is termed Taylor dispersion.
    The dispersion of active suspensions is qualitatively distinct: viscous and external torques can establish non-uniform concentration fields with weighted access to shear, modifying mean drift and effective diffusivity.
    It would be advantageous to fine-tune the dispersion for systems such as bioreactors, where mixing or particle separation can improve efficacy.
    Here, we investigate the dispersion of active suspensions in a vertical channel driven by an oscillatory pressure gradient - Womersley flow - using gyrotactic swimmers (bottom-heavy cells subject to viscous torques).
    Preliminary experimental results reveal interesting dispersion phenomena, highly dependent on the oscillation parameters, motivating theoretical investigation.
    Employing Lagrangian simulations, we find that oscillatory flows can induce drift and increase lateral and downstream dispersion, with periodic mixing between left and right sides. 
    Such flows can also be used to separate species with different motile behaviour.
    Eulerian numerical schemes typically require an approach to averaging in orientational space, such as generalised Taylor dispersion, with assumptions on translational and rotational time scales.
    For an oscillatory timescale commensurate with cell dynamics, we reveal the limitations of such approximations, beyond which the averaging techniques collapse. 
\end{abstract}

% keywords can be removed

\section{Introduction}
Passive tracers in shear flow have a greater effective diffusivity, $D^{e}{'}$, than mass diffusivity alone, due to the combined effect of downstream advection and diffusion across streamlines, a phenomenon known as Taylor dispersion \cite{taylor1953dispersion}. Given a channel or tube of half-width $R'$,mean flow speed $U'$, and mass diffusivity $D'$ it can be shown that $D^{e}{'}=D'(1+\vK \Pe ^2)$, where $\Pe=U' R' / D'$ is the \Peclet number and $\vK$ represents the scale of enhancement and is affected by geometric parameters of the system \cite{taylor1953dispersion,aris1956dispersion}. Replacing passive particles with biased active particles, such as swimming bacteria and microalgae, changes the physical system in a non-trivial manner. For example, some microalgae tend to swim upwards, called gravitaxis, or focus at the centre of downwelling flow in response to a combination of viscous and gravitational/sedimentary torques on the typically bottom-heavy or asymmetric cells, termed gyrotaxis \cite{kessler1986individual}. There are a variety of other taxes that can lead to accumulation of active particles in flow \cite{bees2020advances}. This orientational bias in motility provides dispersion behaviour that is qualitatively distinct from that of tracers: active particles drift relative to the mean flow and their effective diffusivity is strongly dependent on the heterogeneous cell concentration relative to the shear flow gradient \cite{bees2010dispersion}.

Understanding the collective effect of biased motility on dispersion is relevant for systems such as bioreactors that contain active bacteria or the biflagellate microalgae \textit{Chlamydomonas reinhardtii} and \textit{Dunaliella salina} \cite{singh2012development}, utilized to produce materials such as biodiesel and $\beta$-carotene. These organisms are placed in nutrient-rich fluids and may be exposed to light or chemical gradients to drive population growth or illicit particular behaviour. Closed channel bioreactors are preferable to reduce the risk of contamination compared to open-channel designs \cite{zeriouh2017biofouling}, but there is a need for recirculation to prevent accumulation of microorganisms in certain parts of the channel \cite{ahmad2021evolution}. This can be relatively expensive for low value products \cite{borowitzka1999commercial,hoh2016algal}. Theoretical dispersion results are helpful in understanding and eliminating unwanted behaviour, such as biofilm formation \cite{croze2013dispersion}.

Oscillatory flows in channels are ubiquitous in natural environments at a variety of scales, from cardiovascular to tidal systems. Periodic changes of shear direction enhance the diffusion of passive particles \cite{chatwin_longitudinal_1975}, and can be used to separate gases \cite{kurzweg_diffusional_1987} and enhance heat transfer \cite{kurzweg_enhanced_1985}. At low oscillation frequencies, flow profiles are reminiscent of Poiseuille flow, but higher frequencies provide a delay due to finite viscosity \cite{hacioglu_oscillating_2016}, such that the flow near the centre reverses before feeling the impact of the walls from the diffusion of vorticity. Such flows are characterized by the Womersley number, measuring the flow oscillations relative to viscous dissipation. It is given by $\textrm{Wo}=R' \sqrt{\Omega'/\nu'}$, where $\Omega'$ is the angular frequency of the oscillatory flow, $\mu'$ and $\nu'=\mu'/\rho'$ are the dynamic and kinematic viscosities, respectively, and $\rho'$ is the fluid density. Another key parameter is the Schmidt number, $\textrm{Sc}=\nu'/D'$, the ratio of viscosity to mass diffusivity. 

To date, the literature on dispersion in oscillatory flows has focussed exclusively on passive particles \cite{chatwin_longitudinal_1975,purtell_molecular_1981,watson_diffusion_1983}. These studies compute the time-dependent velocity and concentration profiles in various geometries to evaluate the long-time dispersion, and investigate the transient behaviour of the first few oscillations.  Diffusion is enhanced near the wall after just a half-cycle of the flow and the oscillation frequency of the time-varying dispersion is twice the oscillation frequency of the pressure gradient. The studies also report crossover frequencies, in which gases with seemingly different diffusivities are transported at the same rate: geometric parameters and the oscillation frequency can be tuned to maximize dispersion \cite{kurzweg_enhanced_1984,thomas_physics_2001}. Importantly, as the net flow averages to zero, a distribution of passive particles exhibits zero net drift. Recent literature addresses oscillatory dispersion in curved \cite{sharp_dispersion_1991} and annular channels \cite{mazumder_solute_2005}, oscillatory dispersion in a non-Newtonian Casson fluid \cite{nagarani_dispersion_2013}, and oscillatory heat dispersion with leaky boundaries \cite{brereton_diffusive_2017}.

Motivated by the benefits of oscillatory flows on the dispersion of passive particles, it is natural to wonder whether there are useful applications of oscillatory flows in active suspensions as the combination of reorientation and motility in time-dependent shear flows can result in complex trajectories. Measurements of the swimming trajectories of helically swimming active particles (motile \textit{D. salina}) in oscillatory linear shear flows revealed complex resonance effects that can limit swimming progress in the plane of the shear, leading to directed motion \cite{hope2016resonant}. Analysis of the dynamics revealed two important non-dimensional quantities: $\Gamma$, the dimensionless shear rate, and $\Omega$, the driving to intrinsic frequency ratio.  Whilst there were clear resonance effects between the amplitude of the shear via $\Gamma$ and shear-driven Jeffrey orbits for ballistic swimmers, resonance in shear frequency via $\Omega$ was only apparent for helical swimmers, which exhibit an additional intrinsic frequency in their trajectory. Swimming behaviour coupled with more complex flows can lead to a range of ecologically relevant phenomena, such as gyrotactic microswimmers in travelling surface water waves resulting in sub-surface shear trapping \cite{ventrella2023microswimmer}. 

\begin{figure}
    \centering
    \includegraphics[width=0.8\linewidth]{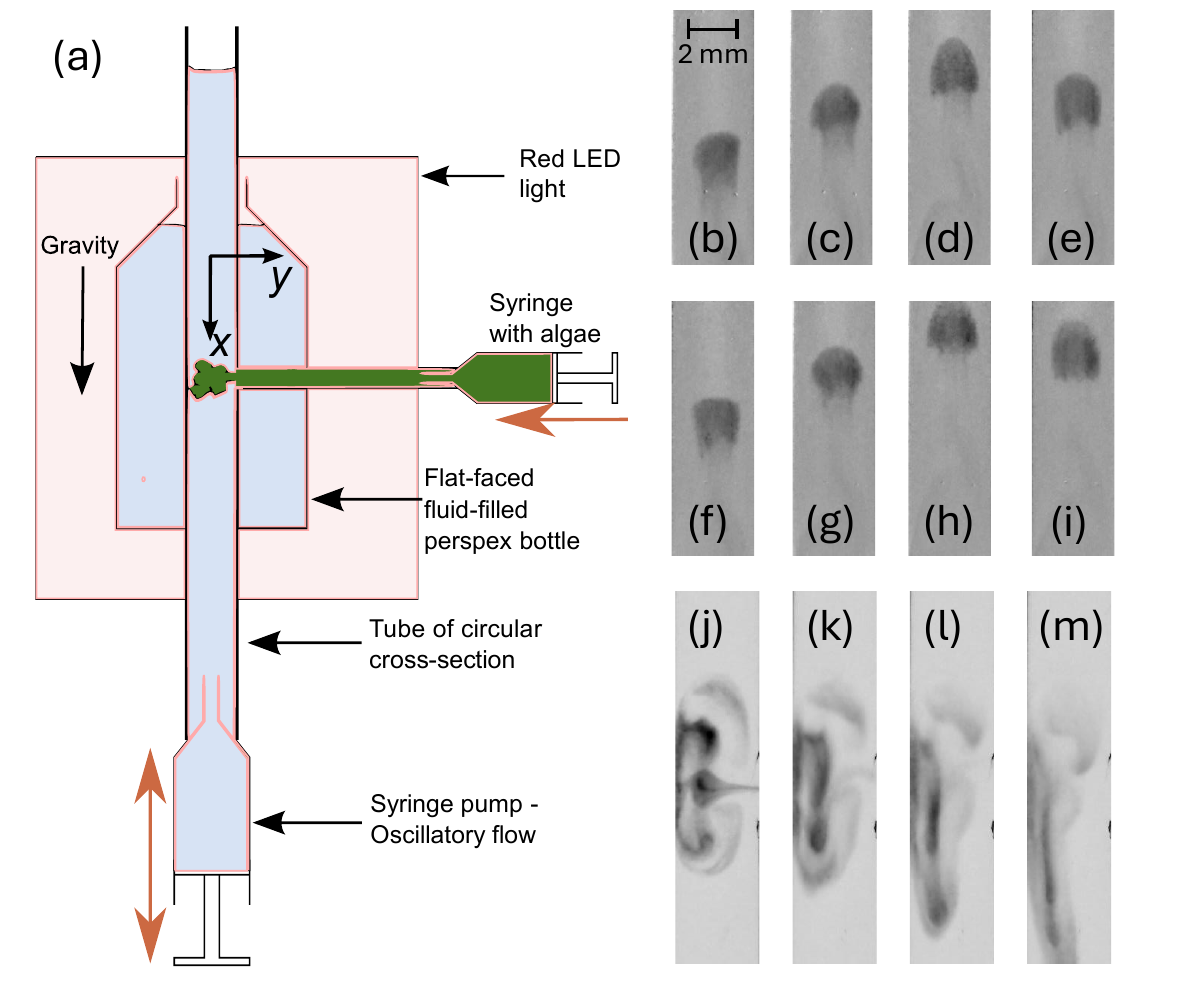}
    \caption{(a) Schematic of the experimental setup (side-view; see main text). (b)-(i) An experiment with Wo=$3.58$ at regular times within two periods of the oscillatory flow, starting before the upwelling flow: (b) $\tau_p=0$, (c) $\tau_p=\pi/2$, (d) $\tau_p=\pi$, (e) $\tau_p=3\pi/2$, (f) $\tau_p=0$, (g) $\tau_p=\pi/2$, (h) $\tau_p=\pi$, and (i) $\tau_p=3\pi/2$. The mean cell position drifts upwards in this instance, with an axisymmetric cell distribution halfway between the tube walls and the central region.  (j)-(m) The transient behaviour for an experiment with a complex initial condition for Wo=$4.52$ at various complete periods of the oscillatory flow: (j) initial distribution, (k) 1 period, (l) 3 periods, and (m) 6 periods.  The cells diffuse in the axial direction, smearing out the initial distribution and focusing cells nearer the centre.  In this case, the mean drift is downwards.}
    \label{fig:exps}
\end{figure}

This paper mostly describes a theoretical study but for motivation we provide some preliminary experimental observations on the impact of upwelling-downwelling oscillatory flows in a channel on the dispersion of a suspension of gyrotactic swimming algae (Fig.~\ref{fig:exps}).  The biased swimming algae respond to the flow, focusing and defocusing, and disperse in a manner qualitatively dissimilar to passive tracers. In particular, the mean position of the cell distribution moves up or down, depending on the driving frequency and amplitude of the flow.  Neutrally buoyant passive tracers diffuse but do not drift.

To establish a theoretical understanding of the system, one can use a continuum or individual-based description of the cells. A complete continuum model would consist of a probabilistic formulation for coupled spatial-orientational aspects \cite{fung_caldag_bees2025}. However, such combined descriptions are at present difficult to deal with.  An alternative and widely used approach to determine the dispersion of active particle suspensions in a tube \cite{bees2010dispersion,bearon2012biased,croze2013dispersion,croze2017gyrotactic}, is to solve an appropriate (phenomenological) advection-diffusion equation for the cell concentration, $n'$, given by
\begin{equation}
    \frac{\partial n'}{\partial t'}= - \nabla \cdot \left[ (\xu' + \Vsp \xq) n' 
    - \xD' \cdot \nabla n'\right],
    \label{eq:n_active}
\end{equation}
\noindent where $\xu'$ is the fluid velocity, $\Vsp$ is the mean swimming velocity, $\mathbf{q}$ is the mean cell orientation, and $\xD'$ is the swimming diffusion tensor.  The first term on the right-hand side describes advection by the flow, the second the mean swimming relative to the flow, and the third the non-isotropic diffusion due to the distribution of cell swimming trajectories in a flow gradient.  Of course, it is then necessary to provide a coherent model for $\xq$ and $\xD'$ as a function of the flow gradient, $\nabla \xu'$ to complete the system.  Generalised Taylor dispersion theory (GTD; Taylor dispersion applied to orientational space) for gyrotactic swimming cells in a linear shear flow \cite{hill2002taylor} has been applied to this end, the solutions of which are used for $\xq$ and $\xD'$ in Eq.~(\ref{eq:n_active}). However, there are significant disadvantages of this approach in the oscillatory flow regime, particularly with regard to assumptions on the relative size of temporal and spatial scales.  
Systems without oscillatory flows are described in terms of a flow time scale $R'/U'$ or swimming time scale $R'^2d_r'/\Vspsq$, where $d_r'$ is the rotational diffusivity. In general for these systems, the swimming time scale is much less than the flow time scale. 
With the oscillating flow, the oscillation time scale is $1/\Omega'$. As we will demonstrate, the GTD approach breaks down if the flow varies on a time scale commensurate with or smaller than that of the reorientation of the cells and/or if spatial flow gradients vary on a scale that is smaller than the spatial scale of cell translation over these times.

Even if it is possible to establish a suitable continuum description (see \cite{fung_caldag_bees2025}) that can deal with the appropriate time-scales and complex interactions of swimmers at boundaries, we face distinct computational issues in attempting to find numerical approximations for the oscillatory dispersion regime except for a limited range of Womersley numbers. Therefore, much of the investigation in this paper attempts to set the groundwork of expectations of the experimental observations using a Lagrangian-based approach, which is much more suited to this problem.  In addition, we probe the regions of applicability and practical limitations of computing solutions for the continuum approach, revealing that low Wo simulations are computationally expensive and large Wo simulations collapse because of the breakdown of the GTD.

The results from this study demonstrate how one can control the axial and lateral dispersion patterns by tuning parameters, particularly the Womersley number. In addition, we find that particles placed exclusively on the left and right halves of the channel can be mixed in an unusual dance in the oscillating flow. Unlike in the case of passive particles where the mean drift is zero, here, we observe non-zero drift directly due to biased motility but in a non-obvious manner. The excess drift can be exploited to separate cells with different motilities. Comparing the solutions for the continuum-based description and individual-based models, we see the descriptions matching for a small range of Womersley numbers: there is an expected breakdown of the continuum approach at large Wo but also computational difficulties arise for fixed \Pe at small Wo.

In the next section, we provide an outline of some preliminary experimental observations, as motivation for the numerical studies.
We then describe the oscillatory flow field and the Lagrangian and Eulerian models, including a novel two-dimensional solution of the GTD description in the Appendix and the various numerical approaches employed.  
The following section describes the results from the numerical simulations, including a demonstration of the relative separation of passive particles and two species with different gyrotactic strengths.
We explain the limitations of the continuum-based solutions through comparisons of results with those from the individual-based model, before reflecting on our results in the discussion.

\section{Preliminary experimental observations}\label{sec:exp_setup}

The experimental setup is displayed in Fig.~\ref{fig:exps}a. Oscillatory flow is generated by a Harvard PHD ULTRA 70-3006 programmable syringe pump, connected via stiff tubing to a plexiglass tube of 7 mm inner diameter. The setup, including the syringe pump, is aligned vertically to generate upwelling-downwelling oscillatory flow. The camera (ELP IF-USB4KCAM30H-CFV) records a region sufficiently far from the ends to ensure a fully developed flow. The tube is placed inside a water-filled reservoir filled to minimise optical aberrations from the edges of the cylindrical tube (similar to \cite{croze2010sheared} for bioconvection in a horizontal tube). Initially, the tube contains media in the absence of algae. The algal suspension is introduced as a blob from an opening at the side of the tube via a second syringe at a concentration of $\sim 10^5$ cells cm$^{-3}$. We use the microalgae \textit{D. salina} from Culture Collection of Algae \& Protozoa (product code CCAP 19/18). The microalgae are cultivated for 2-3 weeks on modified Pick's media under a 16:8 light:dark cycle at 21$^{\circ}$C prior to experiments. A filtering process with a sterile piece of cotton (cells swim up through the cotton) provides control of the cell concentration \cite{croze2017gyrotactic}. To prevent phototaxis, the laboratory is kept dark during the experiments except for a red LED light source (Advanced Illumination) behind the apparatus (Fig.~\ref{fig:exps}a; see \cite{williams2011tale}). The images are captured using MATLAB (image acquisition toolbox) and processed with ImageJ software to enhance contrast. The syringe pump operates with prescribed fluid flux with a periodic square-wave, providing distinct infusion and withdrawal phases. Upon releasing the microalgae, the syringe pump operates for 50 periods. 

Even though the pulsatile flow profile at low Wo should be similar to Poiseuille flow, the full two-way coupled interaction between fluid flow and microswimmer cell concentration can induce flows due to their negative buoyancy, particularly when they accumulate in regions of the flow, and the space-dependent vorticity leads to focusing (see \cite{bees2020advances,wang2023dispersion}). For simple Poiseuille flow imposed fluid pressure determines fluid flux, but for two-way coupled microswimmer suspensions there is a distinct qualitative difference between pressure and flux driven flows allowing access to different solution branches \cite{bees2025emergent}. One might attempt to avoid secondary flow structures by following \cite{croze2017gyrotactic}, fluorescently dyeing some of the negatively buoyant cells in an existing plume structure in a vertical tube, but it is difficult to add dyed cells to a well-developed distribution in this oscillatory case. In this initial study, we inject a dilute suspension of cells into media in a vertical tube and image the cells directly, without the need for dye.

Two examples of experiments are presented in Fig.~\ref{fig:exps}.  The first, in (b)-(i), provides eight images of a carefully-initiated long-time distribution of vigorously swimming cells over two complete flow oscillation periods for Wo$=3.58$ ($R'=3.5$ mm; the properties of the media are similar to water: $\rho' \approx 1050 \text{kg}/\text{m}^3$, $\mu' \approx 0.001 \text{Pa}\cdot\text{s}$; pumping rate of $15$ ml$/$min; pumping/withdrawal volume set to $0.785$ ml). The time instances displayed are expressed in terms of non-dimensional period measure $   \tau_p=\frac{\bmod(t',T')}{T'}2\pi$ where $T'$ is the period of oscillation. Under no-flow conditions with no boundaries, the cells swim upwards on average.  
However, for downwelling (upwelling) flow cells tend to swim towards the centre (to the walls) of the tube in response to gravitational and viscous torques. 
When the flow oscillates with period commensurate with the timescale for reorientation (for this value of Wo, inducing a flatter flow profile) the cells accumulate with a dynamically evolving axisymmetric distribution at an intermediate distance between the centre of the tube and its walls. 
The blob of cells persists but drifts upwards, reminiscent of a swimming jellyfish. In contrast, we observe in Fig.~\ref{fig:exps}(j)-(m) the transient behaviour of cells injected into the flow for Wo $=4.52$. As before, the distribution moves up and down with the flow, but the images in Fig.~\ref{fig:exps} are presented at integer multiples of the flow period. For this slightly larger Wo (larger frequency), the initially complex distribution diffuses and the cells, which are observed to be not as sprightly, descend by focussing into a plume. 
(Videos for both experiments are provided in the Electronic Supplementary Material.) Such qualitative differences in cell distributions and dispersion in response to small flow frequency variations and/or swimming speeds motivate us to explore the system theoretically.

\section{Governing equations}

\subsection{Oscillatory flow field}

Consider a vertically-aligned channel of width $2R'$, with centreline along the $x-$axis and gravity acting in the positive $x-$direction, containing a dilute suspension of non-interacting gyrotactic active particles. Initially, the particles are uniformly distributed across the channel.  The two-dimensional flow field $\uvec'=[u' ~~ v']^T$, where $T$ and $'$ indicate transpose and dimensional quantities, respectively, is governed by the Navier-Stokes equations and driven by an oscillatory pressure gradient $\frac{\difft p'}{\difft x'}$, such that 
\begin{equation}
	\frac{\partial \uvec'}{\partial t'} + (\uvec' \cdot \nabla) \uvec' = \frac{-1}{\rho'} \nabla p' + \nu' \nabla^2 \uvec'
\mbox{~~~~~and~~~~~}
    \frac{\difft p'}{\difft x'} = \Re \{p_a' \exp(i \Omega' t')\}.  \label{eq:dpdx_cos}
\end{equation}
\noindent Here, $p_a'$ is the amplitude and $\Re$ indicates the real part. There is no lateral flow, $v'=0$, and, with a length scale $R'$ and time scale $1/\Omega'$, we can non-dimensionalise the governing equation as
\begin{equation}
    u_{yy}- \frac{R'^2 \Omega'}{\nu'}u_{\tau}=\Re\{\exp(i\tau)\},
\end{equation}
\noindent where $u=u'\mu'/(p_a' R')$, $\tau=\Omega' t'$ and Wo= $\sqrt{ R{'}^{2} \Omega' /\nu'}$. The terms without primes are non-dimensional. No-slip boundary conditions are imposed, $u(\pm 1) = 0$, giving the textbook solution
\begin{equation}
    u(y,\tau)= \Re \left\{ \frac{1}{i \Wo^2} \left( \frac{\cosh (\Wo \sqrt{i} y)}{\cosh (\Wo \sqrt{i})} - 1 \right) \exp(i\tau)   \right\},
    \label{eq:final_flow}
\end{equation}
which diminishes with increasing Wo. In line with the earlier literature, we scale $u$ with a factor $K=1/\langle u \rangle$, where $\langle u \rangle = \sqrt{ 1/(4\pi)\int_{0}^{2\pi}\int_{-1}^{1} u^2\text{d}y\text{d}\tau }$ is the root-mean-square velocity \cite{lee_taylor_2014}, allowing 
a \Peclet number to be defined based on $\langle u \rangle$  
(see Sections 3\ref{sec:lagrangian} and 4\ref{sec:cell_dist}).

\subsection{Lagrangian description}\label{sec:lagrangian}

For individual-based simulations we compute the dynamics for the position $\mathbf{x}_{t'}'$ and orientation $\theta_{t'}$ for each particle $j=0,1,2...N$, where $t'$ indicates the time. The orientation angle $\theta_{t'}$ is defined relative to the $y-$axis, providing the normalized orientation vector $\hat{\mathbf{p}}=[\sin(\theta_{t'}) ~~ \cos(\theta_{t'})]^T$. The particles are assumed to occupy no volume and hydrodynamic interactions are neglected under the dilute assumption. The change in position and orientation of particle $j$, denoted by $\text{d}\mathbf{x'}^j_{t'}$ and $\text{d}\theta^j_{t'}$, are governed by \cite{kessler1986individual,bees2025emergent,croze2013dispersion}
\begin{gather}
    \text{d}\mathbf{x'}^j_{t'} = \mathbf{u'}(\mathbf{x'}^j_{t'},t') \text{d} t'+\Vsp \hat{\mathbf{p}}^j_{t'}\text{d}t',
    \label{eq:sim_x_dim} \\
        \text{d}\theta^j_{t'} = \left(\frac{1}{2B'} \cos(\theta)+\frac{1}{2} \omega'_z(\mathbf{x}'^j_{t'},t') \right)\text{d}t'+\sqrt{2d_r'}\text{d}W_{t'},
    \label{eq:sim_p_dim}
\end{gather}
\noindent where $B'$ is the gyrotactic reorientation time, $\omega_z$ is the $z$-component of the vorticity field and $W_{t'}$ is a Wiener process representing rotational Brownian motion. The amplitude of the noise, $\sqrt{2d_r'}$, ensures a rotational diffusivity of $d_r'$ \cite{croze2013dispersion}.

We follow a cell-based scaling with a time scale $R'^2d'_r/\Vspsq$ and length scale $R'$. In the numerical implementation, $m$ represents the index for time steps with $0 \leq m \leq M-1$, where the dimensionless simulation duration $t_f$ is split into $M$ intervals. Equations~\eqref{eq:sim_x_dim} and \eqref{eq:sim_p_dim} become
\begin{equation}
    \Delta\mathbf{x}^j_m = \Pe_R \mathbf{u}(\mathbf{x}^j_{m},t) \Delt + \beta_R \hat{\mathbf{p}}^j_{m} \Delt,
    \label{eq:sim_x_nondim}
\end{equation}
\begin{equation}
    \Delta\theta^j_m = \left(\frac{1}{2B} \cos(\theta^j_m)+\frac{1}{2} \omega_z(\mathbf{x}^j_{m},t_m) \right)\Delta t+\sqrt{2}\beta_R \Delta W,
    \label{eq:sim_p_nondim}
\end{equation}
\noindent where $\Delta W=W_{m+1}-W_{m}$ represents independent and identically distributed random variables with an expected value of zero and variance of $\Delta t$, following the Euler-Maruyama scheme. The non-dimensional flow and swimming \Peclet numbers are
\begin{equation}
        \Pe_R = \frac{\langle u' \rangle R'd'_r}{\Vspsq} \mbox{~~~~and~~~~} \beta_R = \frac{R'd'_r}{\Vsp},
\end{equation}
respectively, where $\langle u' \rangle$ is the root-mean-square amplitude of the fluid velocity (subscript $R$ indicates $d'_r$-based scaling). The non-dimensional reorientation rate is $B=B'd_r'/\beta_R^2$. One could also follow a flow-based scaling: Appendix \ref{app:scales} describes the relationship between scalings, from which we obtain $\tau=t \text{Wo}^2\text{Sc}$ allowing use of Eq. \eqref{eq:final_flow} in our individual-based model. The particles are confined between $y=\pm 1$. If particle $j$ attempts to cross the boundary, it is specularly reflected \cite{maretvadakethope2023interplay,fung_caldag_bees2025} back to the domain by setting
\begin{gather}
    y^j_{new}=\text{sgn}(y^j) (2-|y^j|) \mbox{~~~~and~~~~}
    \theta^j_{new}=\mbox{mod}(-\theta^j+\pi,2\pi).
\end{gather}

\subsection{Eulerian description}

The governing equation for the non-dimensional concentration, $n$, is written as
\begin{equation}
    \frac{\partial n}{\partial t}=-\nabla \cdot \xJ, \mbox {~~~~where~~~~} \xJ = n\left(\Pe_D \xu + \beta_D \xq\right) - \xD \cdot \nabla n,
    \label{eq:n_nondim}
\end{equation}
where $\xJ$ is the spatial cell flux and we have the non-dimensional numbers
\begin{equation}
    \Pe_D = \frac{\langle u' \rangle R'}{D'}, \quad \beta_D = \frac{\Vsp R'}{D'},
    \label{eq:groups_D}
\end{equation}
based on spatial diffusion scale $D' (= \Vsp^2/d_r')$. No-flux conditions are applied at boundaries:
\begin{equation}
    \mathbf{n}\cdot \left( \beta_D \xq n - \xD \cdot \frac{\partial n}{\partial y} \right)=0 \quad \textrm{on} \quad y=\pm 1,
    \label{eq:nondim_bc}
\end{equation}
\noindent where $\mathbf{n}$ is the boundary unit normal. We follow Bearon et al.'s \cite{bearon2012biased} approach for gyrotactic swimmers, where closed-form expressions are formulated for the diffusion tensor $\xD$ and average swimming direction $\xq$ based on GTD for cells dispersing in a linear shear flow. Here, we derive and use equivalent expressions for $\xD$ and $\xq$ in a two-dimensional channel (see Appendix \ref{app:GTD2}). We prefer $1/\Omega'$ as the time scale to reduce the computational cost. C++ library \texttt{oomph-lib} is employed to solve Eq. \eqref{eq:n_nondim} in an axially periodic box with boundary conditions Eq. \eqref{eq:nondim_bc} \cite{heil2006oomph} based on advection diffusion equations with adaptive Lagrangian quadratic elements and time stepping (as in earlier studies on dispersion of active suspensions \cite{bearon2012biased,maretvadakethope2023interplay,ridgway2023motility}). 
Time derivatives are approximated by a backward difference scheme based on two previous time steps. Fluid flow is introduced as a wind function whilst the swimming of the organisms is represented as a conservative wind function. 
Methods are adapted to accommodate time-dependent shear (impacting swimming and diffusion terms) and flow velocity.

\section{Results}

\subsection{Validation of Lagrangian simulation}\label{}

\begin{figure}
    \centering
    \includegraphics[width=\linewidth]{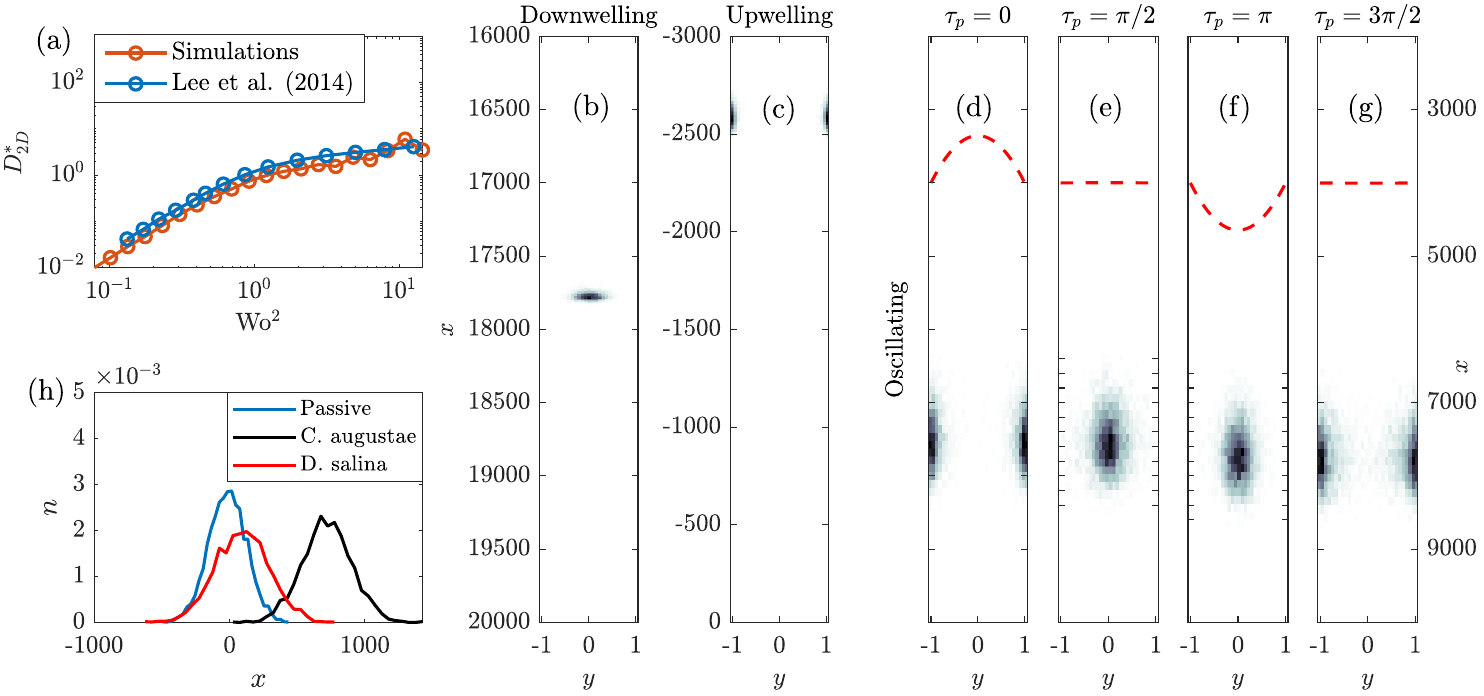}
    \caption{(a) Comparison of individual-based passive particle simulations with results from Lee et al. \cite{lee_taylor_2014}. Concentration distributions of gyrotactic \textit{Chlamydomonas augustae} cells in (b) downwelling, (c) upwelling and (d)-(g) oscillating flows (Wo$=0.106$). (d) is at time $\tau_p=0$, (e) is at time $\tau_p=\pi/2$, (f) is at time $\tau_p=\pi$ and (g) is at time $\tau_p=3\pi/2$. Dashed lines indicate velocity fields at given $\tau_p$ (scaled up and offset for visibility). (h) Long-time axial distributions of passive and gyrotactic particles subject to oscillatory flows, Wo$=0.277$.}
    \label{fig:dists}
\end{figure}

We start by providing a brief validation of our individual-based model with the results from Lee et al. \cite{lee_taylor_2014}, where the dispersion of passive particles under oscillating flows were studied. Following their formulation with a simple spatial diffusivity, the change in position is expressed as
\begin{equation}
    \Delta \mathbf{x}^j_m = \Pe_R \mathbf{u}(\mathbf{x}^j_m,t) \Delt + \sqrt{2D}\Delta\mathbf{W}.
    \label{eq:sim_x_lee}
\end{equation}
\noindent where $D$ is set to unity to compare with Lee et al. \cite{lee_taylor_2014}. We also take $\Pe_R=12.8$ and $\textrm{Sc}=10$ based on the same study.
The measure $D_{2D}^*$ is a period-averaged diffusivity defined by
\begin{equation}
    D_{2D}^*=\frac{1}{2\pi} \int_{\tau_p=0}^{\tau_p=2\pi}  D^{e*} \partial \tau_p,
    \label{eq:D2d}
\end{equation}
\noindent where $D^{e*}$ is the effective axial diffusion, evaluated from variance $V$ in the $x-$direction with
$
    D^{e*}(t)=\frac{1}{2}(V(t)-V(t_m))/(t-t_m),
$
where $t_m$ refers to an end-of-period halfway through the simulation (avoiding transients). If the number of periods is even, $t_m=t_f/2$, otherwise $t_m$ is rounded down to the nearest end-of-period. 
The time span $t-t_m$ is sufficiently long to minimize stochasticity and ensure convergence. The integral in Eq. \eqref{eq:D2d} ranges from $t=t_f-T$ to $t=t_f$ in particle-based scaling. The results in Fig.~\ref{fig:dists}a (varying $\text{Wo}^2$) demonstrate that the individual-based simulations replicate results in \cite{lee_taylor_2014}.

\subsection{Cell distributions}\label{sec:cell_dist}
We begin by examining the effects of biased motility on dispersion with individual-based simulations, based on Eqs. \eqref{eq:sim_x_nondim}-\eqref{eq:sim_p_nondim}. We simulate strongly gyrotactic \textit{C. augustae} (sometimes referred to as \textit{C. nivalis}, see \cite{croze2010sheared}; \textit{C. augustae} has a large $B'$ and small $d_r'$, see Table \ref{tab::simparams} for the simulation parameters).
\begin{table}[t]
\centering
\begin{tabular}{c | c | c | c | c}
\hline
\multicolumn{1}{c|}{Parameter} & $\text{P\'e}_R$                                                                  & Sc                                                                                 & $N$         & $\Delta t$                                                                           \\ \hline
\multicolumn{1}{c|}{Value}     &  1,  \textbf{12.8}, 50                                                                             & 16.8                                                                               & 5000      & \begin{tabular}[c]{@{}c@{}}$T/80$ if Wo $\geq 1$\\  $T/400$ if Wo $< 1$\end{tabular} \\ \hline
Parameter                       & \begin{tabular}[c]{@{}c@{}}$B'$ {[}s{]}\end{tabular}                           & \begin{tabular}[c]{@{}c@{}}$d'_r$ {[}1/s{]}\end{tabular}                         & \begin{tabular}[c]{@{}c@{}}$t_f$ ($\geq$)\end{tabular}   & $\beta_R$                         \\ \hline
Value                           & \begin{tabular}[c]{@{}c@{}} 0.3 for strongly \\ gyrotactic sp.\\ \bf{3.4 for \textit{C. augustae}}\\ 10.5 for \textit{D. salina}\end{tabular} & \begin{tabular}[c]{@{}c@{}}0.01 for strongly \\ gyrotactic sp.\\ \bf{0.067 for \textit{C. augustae}}\\ 0.23 for \textit{D. salina}\end{tabular}   & 1500  & \begin{tabular}[c]{@{}c@{}}0.1 (\textit{C. augustae}) \\ \bf{1 (general)} \\ 5 (\textit{C. augustae})\end{tabular}                                                                                  \\ \hline
\end{tabular}
\caption{Parameters for the individual-based simulations.  Bold indicates base values.}
\label{tab::simparams}
\end{table}
Initially, the particles are uniformly distributed across the channel at $x=0$ with a uniform orientation distribution. In the absence of any flow, the cells would exhibit negative gravitaxis. The flow advects the particles and alters their orientation due to gyrotaxis, resulting in inhomogeneous cross-channel distributions. Long-time distributions in purely downwelling (Fig. \ref{fig:dists}b) and upwelling flows (Fig. \ref{fig:dists}c) are plotted alongside the distributions for an oscillatory flow with Wo$=0.106$ at key $\tau_p$ values (Figs. \ref{fig:dists}d-g). A purely downwelling (upwelling) flow causes migration towards the centre (boundaries) of the channel. Introducing an oscillatory flow field leads to something in-between; the particles move towards the boundaries during upwelling periods but then the flow direction reverses and the cells migrate towards the centre of the channel. Even though the cells swim upwards on average (associated with negative gravitaxis), the particles can drift in the downwards direction on average; the fluid velocity is higher at the centre of the channel and the particles accumulate at the centre of the channel during the downwelling phase. As the particles occupy regions with high shear amplitudes between upwelling and downwelling flows, axial dispersion may also be enhanced. In Fig. \ref{fig:dists}h, we compare the axial distribution of passive and active particles for an oscillatory flow with Wo$=0.277$ at $t=t_f$ (i.e. $\tau_p=0$). The passive particles exhibit a Gaussian distribution centred around $x=0$ whilst the gyrotactic particles drift downwards and generally retain a Gaussian distribution with rate-of-change of variance depending on gyrotactic strength.

\begin{figure}
    \centering
    \includegraphics[width=\linewidth]{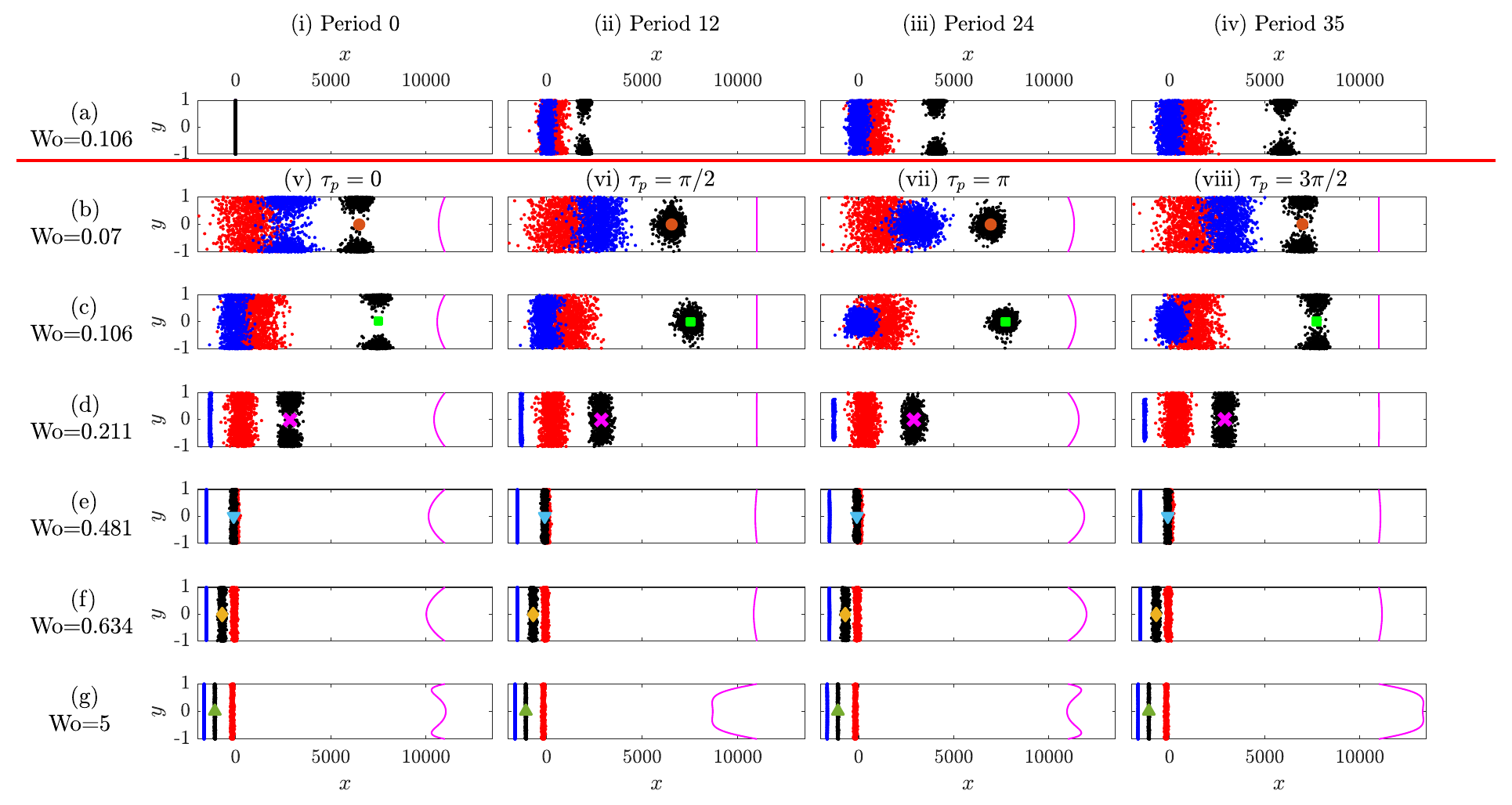}
    \caption{Distributions of \textit{C. augustae} (black), \textit{D. salina} (red) and strongly gyrotactic (blue, see text) particles under oscillatory flow (channels rotated $\pi/2$ anticlockwise). Row (a) gives distributions with Wo$=0.106$ at a given end-of-period, with columns (i)-(iv) indicating periods 0, 12, 24 and 35, respectively. Rows (b)-(g) give the distributions during one period at long times for (a) Wo$=0.07$, (b) Wo$=0.106$, (c) Wo$=0.211$, (d) Wo$=0.481$, (e) Wo$=0.634$ and (f) Wo$=5.00$. Columns (v)-(viii) are for times $\tau_p=0$, $\pi/2$, $\pi$ and $3\pi/2$, respectively. The magenta curves indicate the instantaneous velocity fields ($\times$ $20$; offset by $11000$). Coloured symbols indicate the mean of \textit{C. augustae} and symbols are matched with datapoints in Fig. \ref{fig:drift_disp_cv}.}
    \label{fig:dists_Wo}
\end{figure}

With the focusing mainly governed by the strength of the gyrotactic bias, one can exploit the mechanism to separate cells with different biases. Fig. \ref{fig:dists_Wo}a shows how the cell distributions evolve over time for the same Wo$=0.106$. We plot the distributions for three cell species: weakly gyrotactic \textit{D. salina} (red), strongly gyrotactic \textit{C. augustae} (black) and another model species with even stronger gyrotactic bias for investigative purposes (blue), with motility parameters listed in Table \ref{tab::simparams} (but note that the first two species normally require quite different media). Initial conditions are as above. Over time, the species separate, due in part to how long they occupy the fast-moving regions of the flow. Figs. \ref{fig:dists_Wo}c(v)-(viii) show how the cell distributions change within a period of oscillation at long times. \textit{C. augustae} and strongly gyrotactic cells focus and defocus, sampling the faster-moving part of the fluid and thus drifting away from \textit{D. salina}, which remain broadly spread across the channel. This separation mechanism is non-invasive and it holds great potential to separate active particles from passive particles, or swimming from non-swimming.

The above results set the scene for what we may expect for the impact of upwelling and downwelling phases of the flow on the cells, yet we will see that the results are highly dependent on Wo. Figs. \ref{fig:dists_Wo}b-g show the key moments in a period at long times ($t_f \geq 1500$) for several values of Wo (see Electronic Supplementary Material for an animation of the trajectories for different Wo). The value $\Pe_R=12.8$ is kept constant by scaling $u$ with the factor $K$. At a small value of Wo$=0.07$, the particles except for \textit{D. salina} exhibit separation and regrouping (Figs. \ref{fig:dists_Wo}(b)(i)-(iv)). The focusing at the centre is slightly weaker than that for Wo$=0.106$ (Figs. \ref{fig:dists_Wo}(b)(i)-(iv)). When Wo increases to $0.211$ only strongly gyrotactic cells exhibit focusing, and
they remain spread across the channel 
during the upwelling phase (\ref{fig:dists_Wo}(d)(i)-(iv)). The reduction in particle focusing means that they do not all occupy high-velocity regions, so the mean rate that they drift in the $x$-direction decreases. This decrease is most dramatic for the strongly gyrotactic cells.
The reduced oscillation period also limits axial diffusion.

\subsection{Dispersion characteristics}

\begin{figure}
    \includegraphics[width=\linewidth]{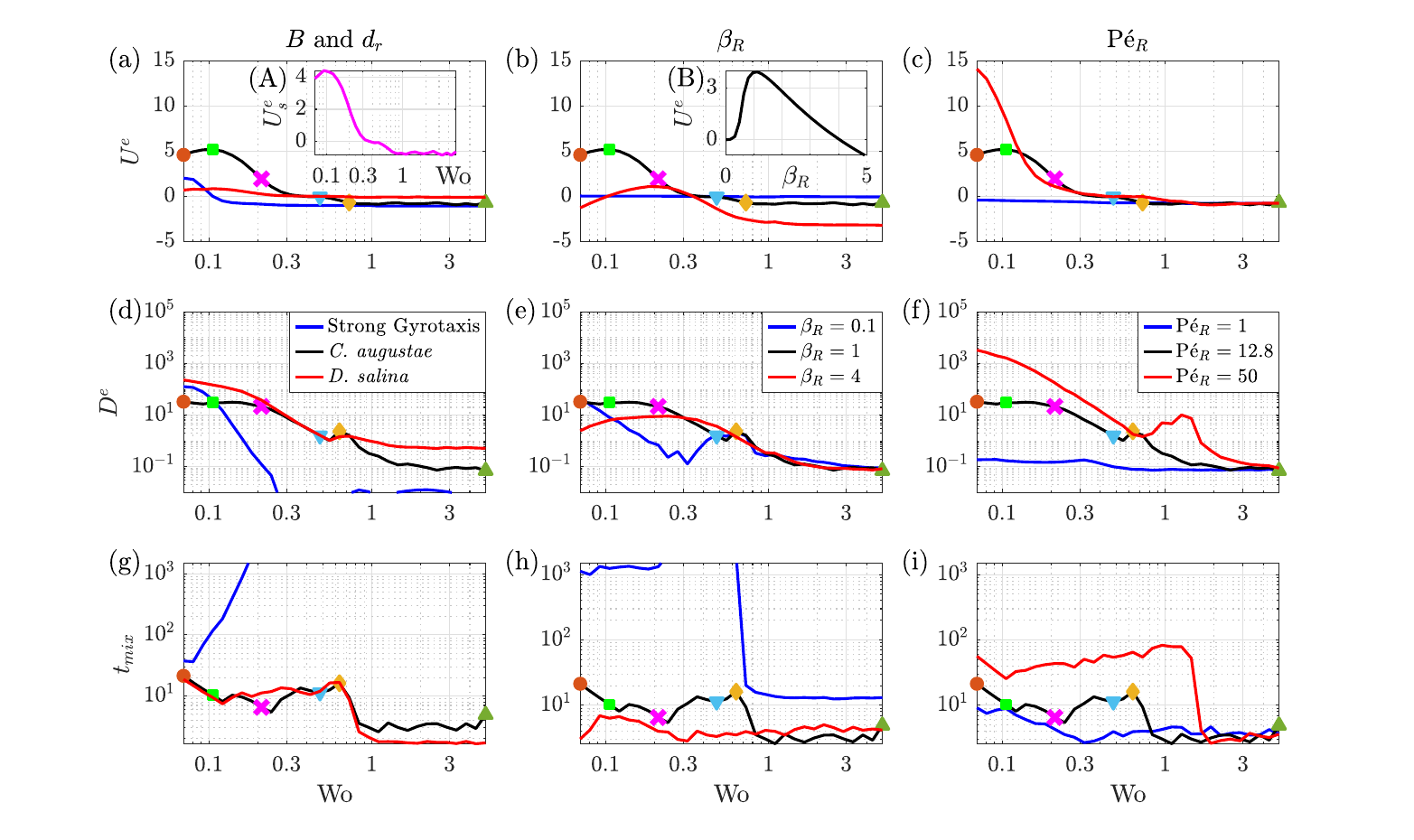}
    \caption{Long-time measures of (a)-(c) excess drift, (d)-(f) axial dispersion and (g)-(i) mixing time (see text) with respect to Wo in individual-based simulations. Plots (a), (d) and (g) show the values for different species (see legend in (d)). Plots (b), (e) and (h) show the values for different $\beta_R$ (see legend in (e)). Plots (c), (f) and (i) show the values for different $\Pe_R$ (see legend in (f)). The black curves in all plots represent the default configuration for \textit{C. augustae} (see Table \ref{tab::simparams} for parameters). The coloured symbols on the black curves indicate critical points and match the cases in Fig. \ref{fig:dists_Wo}. Inset (A) plots the separation velocity between \textit{C. augustae} and \textit{D. salina}. Inset (B) shows the excess drift with respect to $\beta_R$ at Wo$=0.106$ for \textit{C. augustae}.}
    \label{fig:drift_disp_cv}
\end{figure}

The long-time excess drift of the cell distributions is defined by 
    $U^e=\left( x_c(t_f)-x_c(t_l)\right) / \left(t_f-t_l \right)$,
where $x_c$ denotes the axial position of the centre-of-mass of the particles, and $t_l$ and $t_f>t_l$ are long times separated by an integer number of periods, evaluated at the end point of a period to remove within-period fluctuations. Here, we choose five periods to reduce the impact of noise. Figs. \ref{fig:drift_disp_cv}a-c plot $U^e$ with respect to Wo for different cells and $\beta_R$ and $\Pe_R$ values. 
Black curves in Figs. \ref{fig:drift_disp_cv}a-i are for the base configuration for \textit{C. augustae}, see the values in bold in Table \ref{tab::simparams} for the complete set of parameters. Below we summarize our findings:

\begin{itemize}
    \item For \textit{C. augustae} and the range of Wo explored, $U^e$ has a maximum at Wo=$0.106$  (green square in Figs. \ref{fig:drift_disp_cv}a-c, patterns in Fig. \ref{fig:dists_Wo}(b)) but mainly decreases with increasing Wo. The long downwelling phases for small Wo focus the cells at the centre, leading them to drift downwards at the maximum flow rate for half the oscillation period and they accumulate at the walls and do not drift upwards for the other half.
    Strongly gravitactic particles do not accumulate in the centre and at walls as much and, therefore, have smaller drift.
    \item $U^e$ decreases and becomes negative beyond Wo$=0.481$ for \textit{C. augustae} (blue triangles in Figs. \ref{fig:drift_disp_cv}a-c, patterns in Fig. \ref{fig:dists_Wo}(d)). The drift velocity settles around a value of $-0.7$ for Wo$>5.00$ (green triangles in Figs. \ref{fig:drift_disp_cv}a-c, patterns in Fig. \ref{fig:dists_Wo}(f)). As a large Wo number corresponds to fast oscillating plug flow, the net effect of the flow disappears and what remains is the biased (upwards) motility of the particles, gravitaxis.
    \item Fig. \ref{fig:drift_disp_cv}A plots $U^e_s$, the difference between the drift of
    \textit{C. augustae} and \textit{D. salina}. Depending on Wo, \textit{C. augustae} can drift relative to \textit{D. salina} either during the downwelling or upwelling periods. In addition, one can use this mechanism to mix cells that initially are separated. Similarly to the crossover frequencies for gases with different diffusivities\cite{thomas_physics_2001}, here we find $U^e_s\approx 0$ when $\text{Wo}=0.383$.
    \item For small $\beta_R$ (Fig. \ref{fig:drift_disp_cv}b), the particles move relatively slowly, so $U^e$ approaches the zero net advection of the flow. When $\beta_R$ is large, the particles swim faster but with more orientational noise (see Eq. \eqref{eq:sim_p_nondim}). If in addition Wo is small then $U^e$ becomes negative as the cells can swim upwards against the downwelling phase of the flow. As Wo increases, larger fluid velocities lead the cells to focus and drift downwards (positive $U^e$). As before, for even larger Wo, we have a central region of plug flow, the advective effects diminish and gravitaxis dominates leading to upwards drift (negative $U^e$).
    \item Fig. \ref{fig:drift_disp_cv}B shows $U^e$ with respect to $\beta_R$ for Wo$=0.106$. The drift peaks at around $\beta_R=1$, where the swimming and rotational diffusion timescales coincide, providing maximal cell focusing at the centre of the channel during the downwelling period.
    \item The response to $\Pe_R$ is displayed in Fig. \ref{fig:drift_disp_cv}c. The particles swim upwards at low $\Pe_R$ against the weak flow while at large $\Pe_R$ the particles exhibit strong downwards drift. As the flow effects diminish at large Wo, the curves for different $\Pe_R$ collapse.
\end{itemize}

The effective axial diffusion $D^{e}=D^{e*}(t_f)$ is plotted in Figs.~\ref{fig:drift_disp_cv}d-f. In summary:
\begin{itemize}
    \item For \textit{C. augustae}, $D^{e}$ is greatest for Wo=$0.07$ (orange circle in Fig.~\ref{fig:drift_disp_cv}d, patterns in Fig. \ref{fig:dists_Wo}(a)), which then decays to zero as Wo increases due to a flattening of the distribution across the channel with smaller shear sampled more often, except at the walls.
    \item There is a slight increase in $D^{e}$ that coincides with $U^e$ going negative (golden diamonds in Figs. \ref{fig:drift_disp_cv}a and \ref{fig:drift_disp_cv}d, patterns in Fig. \ref{fig:dists_Wo}(e)), which is where we start to see a qualitative change in the shear profile.
    \item $D^e$ for \textit{D. salina} follows that of \textit{C. augustae} at intermediate Wo where both species are spread across the channel. At large Wo, the relatively weak gyrotaxis of \textit{D. salina} induces greater $D^e$. For small Wo, \textit{D. salina} are still spread across the channel but \textit{C. augustae} repeatedly focus and separate, spending extended periods in central regions of relatively low shear, yielding a smaller $D^e$.
    \item For strong gravitaxis, the cells do not diffuse much across streamlines and thus $D^e$ is relatively small (and numerically noisy). 
    For small Wo the cells have time to slowly transit the high shear regions leading to relatively large values of $D^e$ (see \ref{fig:drift_disp_cv}c).
    \item $D^e$ is insensitive to $\beta_R$ at large Wo (Fig. \ref{fig:drift_disp_cv}e) as the flow profile is mostly flat across the channel, but for small Wo there is an interesting dependence on $\beta_R$ for the balance between focusing and swimming across high shear regions (see the blue curve in particular).
    \item Fig. \ref{fig:drift_disp_cv}f displays $D^e$ for different $\Pe_R$. For small $\Pe_R$, $D^e$ is mostly flat and for large $\Pe_R$ we see an approximate $\text{Wo}^{-4}$ dependence for small Wo. There is also a local maximum for $\Pe_R=50$ at Wo $\sim$ 1 associated with transit across high shear regions. 
\end{itemize}

Lastly, we explore cross-channel mixing through the quantity $t_{mix}$, a measure based on the signal-to-noise ratio, $\textrm{SNR}={\mu}/{\sigma}$, where $\mu$ is the mean and $\sigma$ is the standard deviation of $y^j$ for particles initially placed on the right half of the vertical channel (i.e. $y^j_0>0$). If these particles spread across the channel, $\mu$ and $\sigma$ should approach zero and 0.58, respectively, and the SNR will tend to zero. The time it takes for the SNR to fall below a reference value of $0.1$ is called $t_{mix}$. Figs.~\ref{fig:drift_disp_cv}g-i plot $t_{mix}$ values for all simulations.  In summary:

\begin{itemize}
    \item For \textit{C. augustae}, $t_{mix}$ is large at small Wo. The distinct minimum at Wo$=0.211$ (magenta cross in Figs.~\ref{fig:drift_disp_cv}g-i, patterns in Fig. \ref{fig:dists_Wo}(d)), corresponds to the smallest Wo for which the particles are spread across the whole channel. Beyond this Wo, $t_{mix}$ increases until Wo=$0.634$ (golden diamonds in Figs. \ref{fig:drift_disp_cv}g-i, patterns shown in Fig. \ref{fig:dists_Wo}(f), coinciding with the point of the local maximum in $D^{e}$ and the change in sign of $U^e$.
    \item For both \textit{C. augustae} and \textit{D. salina}, we see a sharp decrease in $t_{mix}$ around Wo$=1$ and the values remain low for Wo$>1$. The decrease in oscillation period, coupled with the change in the flow profile and thus trajectories of the active particles, clearly enhances mixing.
    \item $t_{mix}$ for \textit{D. salina} is similar to that of \textit{C. augustae} for small Wo, yet the value is significantly smaller for large Wo due to reduced biased motility. When gravitaxis is strong, mixing times are very long.
    \item Small $\beta_R$ delays mixing due to slower swimming (Fig. \ref{fig:drift_disp_cv}h). Large $\beta_R$ improves mixing to the extent that the sharp drop in $t_{mix}$ around Wo$=1$ can be removed.
    \item Large $\Pe_R$ flows delay the mixing by suppressing the diffusion and biased motility of the cells (Fig. \ref{fig:drift_disp_cv}i).
\end{itemize}

\subsection{Limits of validity of the Eulerian description for time-varying flows}

For the Eulerian simulations, we attempt to compute the concentration field $n$ for gyrotactic active particles. The domain length for simulations is set to 700 and the initial distribution of particles is $n(x,t=0)=e^{-0.01\left|  x-x_c^0 \right|^2},$ where $x_c^0$ represents the coordinate of the peak of the distribution, selected based on the drift of the blob of particles to minimize end effects. There are numerous computational challenges associated with the Eulerian approach. One cannot explore parameter space easily for very small Wo due to the excessive movement of the distribution for that case. This necessitates large computational domains to prevent the concentration profile sitting across the streamwise periodic structure and interacting with itself. Furthermore, long simulation times are required to obtain (periodically) converged solutions. Due to the high computational cost of the numerical model, we are only able to simulate a non-dimensional duration of $t=300$ in a reasonable time, compared to the $t=1500$ runs accessed with the Lagrangian approach. In addition, we set $\Pe_R=2$ to prevent overlap within a domain of manageable size. We use $\mathbf{q}$ and $\bm{D}$ obtained for \textit{C. augustae} (see Appendix \ref{app:GTD2}) and Lagrangian simulations use the corresponding $d_r'$ and $B'$ values. Lastly, $\beta_R=1$ for all simulations reported in this section.

\begin{figure}
    \centering
    \includegraphics[width=0.8\linewidth]{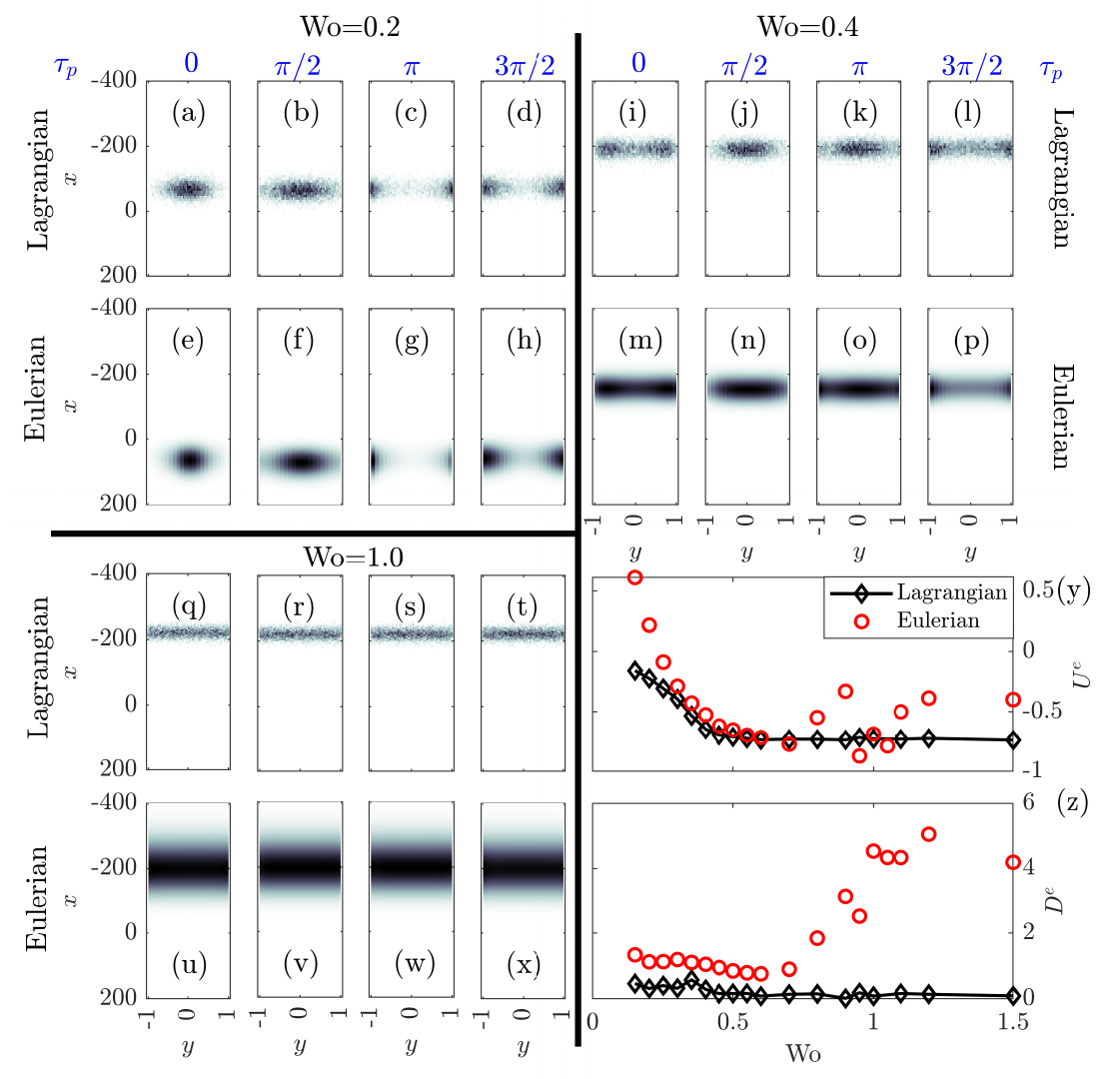}
    \caption{Comparison between Lagrangian and Eulerian model results. (a)-(x) show $n$ as density plots, with darker colors indicating larger $n$. Columns represent different $\tau_p$ with the values shown at the top of each column. (a)-(h) show the results for Wo$=0.2$, (i)-(p) are for Wo$=0.4$ and (q)-(x) are for Wo$=1.0$. (a)-(d), (i)-(l) and (q)-(t) show Lagrangian simulation results while (e)-(h), (m)-(p) and (u)-(x) show Eulerian simulations. (y) and (z) compare $U^e$ and $D^e$ obtained from these models, respectively.}
    \label{fig:cont_sims}
\end{figure}

Fig.~\ref{fig:cont_sims} compares the concentration field $n$ obtained via the Eulerian and Lagrangian schemes (See Electronic Supplementary Material for an animation.). Three values of Wo are selected to help explain similarities and differences in the results of the approaches. For Wo$=0.2$ (Figs. \ref{fig:cont_sims}a-h), cross-channel distributions of particles are remarkably similar, showing the focusing and separation discussed earlier. However, whilst the Lagrangian particles show clear cell migration against gravity, the drift is downwards in the Eulerian simulations (Fig.~\ref{fig:cont_sims}y). This $\Pe_R$ value (in contrast to $\Pe_R=12.8$ in Fig.~\ref{fig:dists_Wo}) represents a regime where the gravitaxis of the particles remains slightly stronger than the downwards fluid flow. In the Eulerian simulations, as particles migrate to the boundaries, they experience a changing shear flow.  The shear may be sufficiently strong that they tumble, yielding a mean swimming vector $\mathbf{q}$ with magnitude much smaller than one. On the other hand, $\mathbf{\hat{p}}$ and $\mathbf{x}$ in the Lagrangian simulations change dynamically, providing complex swimmer trajectories, which may cross the centre or intersect with boundaries.  The no-flux condition used in the Eulerian simulations approximates this boundary interaction, losing information about prior trajectories \cite{fung_caldag_bees2025}. The specularly reflective boundary conditions employed in the Lagrangian simulations are a better representation of the real boundary interactions and can significantly effect the cell distribution \cite{maretvadakethope2023interplay}, and thus dispersion characteristics, during the upwelling part of the flow.

As Wo increases, in an intermediate region, the two descriptions show closer agreement. Figs. \ref{fig:cont_sims}i-p present distributions for Wo$=0.4$. Both the distributions and $U^e$ match better for this value of Wo, as seen in Fig. \ref{fig:cont_sims}y. However, note that $D^e$ from the Eulerian simulations are always larger than those from the Lagrangian simulations. Stronger and more persistent accumulation around the boundary and the quasi-steady nature of the mean orientational dynamics allow particles more time in the high-shear regions and this leads to a notably larger $D^e$. 

As Wo is increased beyond $0.7$, we once again observe strong divergence of the results from each other. Figs.~\ref{fig:cont_sims}q-x provide distributions for Wo$=1.0$. While $U^e$ values are remarkably close, the irregularity of $U^e$ at other Wo values in Fig. \ref{fig:cont_sims}y indicates that this is more of a coincidence. 

The quasi-steady state GTD expressions for $\mathbf{q}$ and $\bm{D}$ used in the Eulerian simulations are evaluated for a linear shear flow.  If the cells traverse a flow field that changes in space or time faster than the cell orientation dynamics then we are unlikely to obtain consistent results.
Indeed, the Eulerian simulations, and particularly the GTD approach, are expected to break down when the cell reorientation timescale $R'^2 d_r'/\Vsp^2$ is larger than the flow oscillation timescale $1/\Omega'$, which is when Wo$^2$Sc $\gtrsim 1$, and we observe results diverge with increasing Wo number starting from a value somewhat less than one.  
However, the breakdown mostly occurs within the high shear regions of width $1/\mbox{Wo}$ close to the walls, so whilst the breakdown should not affect dynamics within the relatively uninteresting plug flow region it does make the overall Taylor dispersion results suspect in the Eulerian simulations for large Wo.  In the large Wo limit, cells will be confined to the broad plug flow region and without a shear flow the oscillation time scale $1/\Omega'$ is no longer relevant; for Wo$\gg 1$ cells swim upwards with the same bias as for no fluid flow, providing a drift of $-I_0(\lambda)/I_1(\lambda)$ (see Appendix \ref{app:GTD2}), which for $\lambda=2.2$ is -0.7281, once again agreeing with the results from the Lagrangian simulation.  

\section{Conclusion}

Active suspensions exhibit a rich variety of new phenomena when they are subject to fluid flows.  
In particular, active particles disperse in a flow in a tube in a manner that is qualitatively distinct from that of passive tracers \cite{bees2010dispersion,croze2017gyrotactic}.  
This is caused by the complex interaction between advection and individual particle dynamics, driven by fluid shear.
Previous analyses have restricted attention to steady flows, as steps to coarse-grain the underlying microscopic individual-based models to a macroscopic continuum model typically assume that the time scales of active particle dynamics are much smaller than variations in the flow \cite{fung_caldag_bees2025}.
It is unclear whether the resulting phenomenological advection-diffusion equations for the cell concentration are adequate in situations where the flow field varies over a time scale commensurate with the cell reorientation.

This study is motivated by preliminary experimental observations, reported herein, that investigate the behaviour of active suspensions under oscillatory flows in a tube; we subject a suspension of gyrotactic swimming cells to upwelling-downwelling oscillatory flow.
For steady Poiseuille flow, cells would either migrate to the walls or towards the centre of the channel, depending on whether the flow is upwelling or downwelling, respectively. 
However, experiments suggest that oscillating flows enable a range of behaviour, with various amounts of drift (in contrast to passive particles) and effective diffusion realised as a function of the system parameters, particularly the Womesley number, Wo.  

We use individual-based simulations to show how the active particles migrate from one distribution to the other as the flow switches direction. 
At intermediate oscillation frequencies, the particles exhibit complex patterns between these distributions, dispersing axially and laterally in the channel.  
The shear experienced by the particles, and their resulting trajectories, enhances the mixing of particles placed exclusively on either side of the channel. 
Furthermore, the gyrotactic particles exhibit an excess drift in oscillatory flows.  
We propose a new method of separating two algal species with different motilities by exploiting their relative drift in oscillatory flows.

The Lagrangian simulations are contrasted with a finite-element based approximation of the phenomenological Eulerian description of the system. The Eulerian approach utilizes a mean swimming direction, $\xq$, and diffusion tensor, $\xD$, computed from generalised Taylor dispersion theory as a function of vorticity. We provide novel closed-form approximations in two-dimensions for this purpose.  The results of the continuum-based model agree with the individual-based description for a small range of Womersley numbers. However, the current Eulerian approach breaks down for large Wo as the oscillation time scale approaches the diffusive time scale and invalidates the quasi-steady state assumption. At low Wo, the approach is computationally impractical and factors such as boundary conditions and particle behaviour at large Wo cause a divergence from Lagrangian based results. Clearly, for the problem of dispersion in oscillatory flows, the individual-based simulations are more accurate and much preferred computationally. To access approximations at large Wo, a complete solution of the full Smoluchowski equation is required, describing the probability, $P(\xx,\xp,t)$, that there is a particle at position $\xx$ with orientation $\xp$ at time $t$ 
\cite {hill2002taylor,fung_caldag_bees2025}, given initial data. This, however, presents considerable computational hurdles of its own.

The relative impact of oscillatory flows on the dispersion of active particle suspensions presents a simple yet powerful method for a variety of biological and medical applications. For example, the controlled dispersion of active particles can be useful in bioreactors as a passive mixing mechanism away from surfaces, reducing the potential for biofilm formation. The oscillatory flows can also provide demixing for cell separation. Our approach is non-invasive and it could be developed into a fully-closed cost-effective and practical system. Finally, we hope that our findings motivate the development of new simplified continuum descriptions of active suspensions that are robust to rapid variations in flow fields.

\appendix
\section{Appendix}
\subsection{Scaling considerations}\label{app:scales}

The system analysed in this manuscript permits different scalings. The oscillating flow field solution provides the Womersley number Wo=$R'\sqrt{\Omega'/\nu'}$. Here, $R'$ and $1/\Omega'$ are the length and time scales, respectively. The non-dimensional time in terms of this scaling is $\tau=t' \Omega'$.  Eqs.~\eqref{eq:sim_x_dim} and \eqref{eq:sim_p_dim} become

\begin{equation}
    \Delta\mathbf{x}^j_m = \Pe_F \mathbf{u}(\mathbf{x}^j_{m},t) \Delta \tau + \beta_F \hat{\mathbf{p}}^j_{m} \Delta \tau,
\end{equation}
\begin{equation}
    \Delta \theta^j_m= \left(\frac{1}{2B} \cos(\theta^j_m)+\frac{1}{2} \omega_z(\mathbf{x}^j_{m},t_m) \right)\Delta \tau+\sqrt{2d_r}\Delta{W},
\end{equation}
\noindent where $\Delta {W}$ has variance $\Delta \tau$. The non-dimensional numbers now become
\begin{equation}
    \Pe_F=\frac{\langle u' \rangle}{R' \Omega'}, \quad \beta_F=\frac{\Vsp}{R' \Omega'},
\end{equation}
\noindent where the subscript $F$ indicates flow-based scaling. The non-dimensional reorientation rate is $B=B'\textrm{Wo}^2\textrm{Sc}d_r'$ and the non-dimensional reorientation rate is $d_r=1/\textrm{Wo}^2\textrm{Sc}$. The relations between the cell and flow-based \Peclet numbers are
\begin{equation}
    \Pe_R=\Pe_F \textrm{Wo}^2\textrm{Sc}, \quad \beta_R=\beta_F \textrm{Wo}^2\textrm{Sc}. 
    \label{eq:cell_to_flow}
\end{equation}
In the cell-based scaling $t=t' \Vspsq/(R'^2d'_r)$, where $D \sim \Vspsq/d'_r$. Defining the Schmidt number as Sc$=\nu'd'_r/\Vspsq$, we can express $\tau=t \textrm{Wo}^2\textrm{Sc}$.

\subsection{Two-dimensional generalised Taylor dispersion for gyrotactic micro-organisms in linear shear flow} \label{app:GTD2}

The approaches of \cite{hill2002taylor} and \cite{bearon2012biased} employ a Galerkin method to determine expressions for the mean swimming direction, $\xq$, and diffusion tensor, $\xD$, in a steady linear flow using generalised Taylor dispersion theory (GTD) for the phenomenological advection-diffusion equation in Eq.~(\ref{eq:n_nondim}) for the cell concentration, $n$.  In particular, they consider three-dimensional physical space with cell orientation determined by two Euler angles. The dimensional micro-scale model is 
\begin{equation}
	\dtp{P} + \nabla_{\xx}\cdot \left[ \left( \xu + \Vsp \xp \right) P\right] + \cL P = 0,  \mbox{~~~~~where~~~~~} \cL = \nabla_{\xp} \cdot \left[ \dot{\xp} - d_r \nabla_{\xp}  \right], 
\end{equation}
for $P(\xp, \xx ,t)$, the probability of there being a cell at $\xx$ with orientation $\xp$ at time $t$, 
where
$\dot{\xp}$ can be evaluated from Eq.~(\ref{eq:sim_p_dim}).  By considering the long-time limits of the co-deformational derivatives of the moments of the distribution $P$, GTD then provides expressions for mean cell orientation, $\xq$, and diffusion tensor, $\xD$, as integrals over cell orientation, $\xp$, in the form 
\begin{equation}
	\xq = \int_{\xp} \xp f(\xp) d\xp \mbox{~~~~~and~~~~~} \xD = \int_{\xp} \left[ \xb \xp +\frac{2\sigma}{f(\xp)} \xb \xb \cdot \hG   \right]^{\mbox{sym}} d\xp, \label{eq:qnD}
\end{equation}
where $\sigma = \Pe/2 \chi'$, with $\chi'$ indicating derivative of the flow field relative to the mean (assuming $\Pe$ is defined with respect to the mean). $[]^{\mbox{sym}}$ indicates the symmetric part and $\hG$ is the transpose of the non-dimensional fluid velocity gradient, whose eigenvalues must be imaginary for the method to work. To evaluate these expressions one must find solutions for the equilibrium and vector field distributions, $f(\xp)$ and $\xb(\xp)$, respectively, via 
\begin{equation}
	\cL f = 0 \mbox{~~~~~and~~~~~} \cL \xb -2\sigma \xb \cdot \hG = f(\xp) (\xp -\xq),
\end{equation}
with $\int_{\xp} f(\xp)d\xp = 1$ and  $\int_{\xp} \xb(\xp)d\xp = \mathbf{0}$,
as described in \cite{hill2002taylor}.  Solutions are expanded in spherical harmonics, substituted into the equations, and truncated at a particular order.  Coefficients are determined using exact algebra (Maple) as a function of $\lambda$ and $\sigma$, with excellent agreement with asymptotic limits for small and large $\sigma$.

In this article, we mirror this approach but consider all variables constrained to a two-dimensional plane, such that 
$\xx=[x ~~ y]^T$ and $\xp = [-\cos \theta ~~ \sin \theta]^T$, $\theta \in [0,2\pi)$.  The variable $x$ is measured in the vertical downwards direction, $y$ is across the channel and $\theta$ is the angle from the positive $x$-axis.  In this situation, we find that $\dot{\theta}=-\frac{1}{2B}\sin\theta -\frac{G}{2}$ and thus we need to solve 
\begin{equation}
	\cM f = 0, \mbox{~~~~~} \cM b_2 = -f (\sin\theta -q_2)\mbox{~~~~~and~~~~~}\cM b_1 -2\sigma b_2 = f (\cos\theta +q_1), \label{eq:fnb}
\end{equation}
most conveniently accomplished in that order, 
where the operator $\cM$ is given by
\begin{equation}
	\cM = \dtt{} +(\lambda \sin \theta +\sigma) \dth{} +\lambda\cos\theta.
\end{equation}
Expansion in Fourier series of the form $C_i = \sum_{n=0}^{\infty} A_n^i\sin n\theta +B_n^i \cos n\theta$, where $C^0=f$, $C^1=b_1$ and $C^2=b_2$, and truncating at a suitable order provides a system of algebraic equations to solve for the coefficients.   Substituting the series in Eq.~(\ref{eq:qnD}) provides the expressions 
\begin{equation}
	\xq = \pi[-B_1^0 \quad A_1^0]^T \mbox{~~~~~and~~~~~} \xD = \frac{\pi}{2}\left[ 
	\begin{array}{cc}
		-2B^1_1 & A^1_1 - B^2_1 \\ 
		A^1_1 - B^2_1 & 2A^2_1
	\end{array}  
\right]  - \int_{\xp}  \frac{2\sigma}{f(\xp)} \left[ \begin{array}{cc}
	2b_1 b_2 & b_2 b_2 \\ 
	b_2 b_2 &0
\end{array} \right]  d\xp.
\end{equation}
Solutions converge rapidly; an order three truncation is suitable for most purposes. Similar to \cite{bearon2012biased}, the solutions are well-fitted by the curves of the form $
	q_1 = - F(\sigma; \xa^1,\xc^1),\mbox{~~~}q_2 = -\sigma F(\sigma; \xa^2,\xc^2),
$
and similarly for odd and even components of the diffusion tensor, where 
$	F(\sigma; \xa,\xc) = (a_0+a_2\sigma^2)/(1+c_2\sigma^2+c_4\sigma^4)$. Coefficients for $\lambda=2.2$ are provided in Table \ref{tab::continuum_coef}.

\begin{table}[]
\centering
\begin{tabular}{c|c|c|c|c}
\hline
Term & $a_0$                   & $a_2$                                & $c_2$                  & $c_4$                  \\ \hline
$q_1$     & $7.281 \times 10^{-1}$  & $7.52 \times 10^{-2}$                       & $3.584 \times 10^{-1}$ & $6.06 \times 10^{-2}$  \\ 
$q_2$     & $3.814 \times 10^{-1}$   & $4.489 \times 10^{-2}$                        & $2.959 \times 10^{-1}$ & $4.405 \times 10^{-2}$ \\ 
$D^{11}$  & $6.9693 \times 10^{-2}$ & $3.046 \times 10^{-1}$                        & $1.627 \times 10^{-1}$                & $8.72 \times 10^{-2}$                \\ 
$D_{22}$  & $1.7672 \times 10^{-1}$ & $1.269$ & $7.959$ & $1.483$ \\ 
$D_{12}$  & $1.694 \times 10^{-1}$  & $5.182 \times 10^{-2}$                        & $2.166 \times 10^{-1}$ & $9.147 \times 10^{-2}$ \\ \hline
\end{tabular}
\caption{The fit coefficients for $\lambda=2.2$ for the mean swimming direction and diffusion.}
\label{tab::continuum_coef}
\end{table}

To verify the method, solutions can be found for $\sigma=0$ for the two-dimensional case.  Eq.~(\ref{eq:fnb}) provides $[\lambda(1-x^2)^{\frac12} (f-f')]'=0$, upon substitution of $x=\cos\theta$.  Integrate and determine the constant at $x=1$ to be zero, to yield the von Mises distribution $f(\theta)=A e^{\lambda \cos\theta}$.  Noting that modified Bessel functions of the first kind can be defined as $I_n(\lambda)=\frac{1}{2\pi} \int_0^{2\pi}  e^{\lambda \cos\theta} \cos n\theta ~d\theta$, we have $A=1/2\pi I_0(\lambda)$, from which we can calculate $\xq = [-\frac{I_1(\lambda)}{I_0(\lambda)}  \quad 0]^T.$  For $\lambda=2.2$, $q_1 = -0.7281$, which agrees with the series solution, as does the distribution for $f(\theta)$.  Note that for the three dimensional case the equivalent $q_1=-\coth \lambda + 1/\lambda=-0.57$, so the two-dimensional case provides a larger peak orientation, as one may expect, plus it also decays to zero a little faster and the $y$ component has a turning point at a smaller $\sigma$ (by a factor of $2/3$; the cell will tumble for smaller shear rates).  Qualitatively, the solutions are similar to the three-dimensional case.

Similarly, we can integrate the equation for $b_2$ to find
\begin{equation}
	b_2 = \frac{e^{\lambda \cos\theta}}{2\pi I_0(\lambda)\lambda}    
	\left( \theta - \frac{1}{I_0(\lambda)}   \int_0^{\theta} e^{-\lambda \cos\theta'} d\theta'  \right).
\end{equation}
A more involved expression can be found for $b_1$, from which one may compute components of the diffusion tensor numerically, establishing excellent agreement with the earlier solution from the series expansion ($D_{11}=0.06969$, $D_{12}=0$ and $D_{22}=0.1767$).  One may also check that the eigenvalues of the diffusion tensor remain positive definite for all $\lambda$ and $\sigma$. Diffusion values can also be fitted by curves, with $D_{11}=F(\sigma; \xa^{11},\xc)$, $D_{12}=-\sigma F(\sigma; \xa^{12},\xc)$ and $D_{22}=F(\sigma; \xa^{22},\xc)$.

%\bibliographystyle{RS}
%\bibliography{Bib}  %%% Remove comment to use the external .bib file (using bibtex).

\printbibliography

\end{document}